\documentclass{article}
\usepackage{spconf,amsmath,graphicx}

\usepackage{booktabs, hyperref}
\usepackage[dvipsnames]{xcolor}
\renewcommand{\paragraph}[1]{\noindent\textbf{#1}}

\title{Cross-lingual Transfer for Speech Processing \\
using Acoustic Language Similarity}
\name{Peter Wu, Jiatong Shi, Yifan Zhong, Shinji Watanabe, Alan W Black}
\address{Carnegie Mellon University, PA, USA\\
\tt{peterw1@cs.cmu.edu}}
\begin{document}
\ninept
\maketitle
\begin{abstract}
Speech processing systems currently do not support the vast majority of languages, in part due to the lack of data in low-resource languages. Cross-lingual transfer offers a compelling way to help bridge this digital divide by incorporating high-resource data into low-resource systems. Current cross-lingual algorithms have shown success in text-based tasks and speech-related tasks over some low-resource languages. However, scaling up speech systems to support hundreds of low-resource languages remains unsolved. To help bridge this gap, we propose a language similarity approach that can efficiently identify acoustic cross-lingual transfer pairs across hundreds of languages. We demonstrate the effectiveness of our approach in language family classification, speech recognition, and speech synthesis tasks.
\end{abstract}
%
\section{Introduction}
\label{sec:intro}

As speech assistants and other speech-based technologies become more prevalent, it is important to make them accessible to everyone around the world. While there are thousands of languages in the world, these technologies currently only support a small subset of these languages. Since state-of-the-art speech processing algorithms typically require large training corpora, it is difficult, if not infeasible, to currently build them for low resource languages\cite{wang2017tacotron, rallabandi2019submission}.

Cross-lingual transfer is a promising direction that aims to bridge this gap by incorporating high-resource data into low-resource systems \cite{dalmia2018sequence, hou2020large}. For such approaches, choosing the data to transfer from is an important step \cite{conneau2020xlsr, li2019corpus, chen2018multi, adams2019massively_shinji_asr}. While many solutions for this step exist for text-based tasks, generalizable solutions still do not exist for speech-based ones \cite{lin-graham-2019-choosing}. In this work, we help bridge this gap by proposing language similarity metrics that pair target languages with source languages suitable for acoustic cross-lingual transfer. We find our approach effective in language family classification, speech recognition, and speech synthesis. Thus, our contributions are the following:
\begin{enumerate}
    \item we introduce an acoustic language similarity approach,
    \item we compare it with non-acoustic ones in downstream cross-lingual speech recognition and speech synthesis tasks, and
    \item we demonstrate the usefulness of our acoustic approach for both language family classification and acoustic cross-lingual transfer.
\end{enumerate}

We proceed by discussing cross-lingual transfer techniques in Section~\ref{sec:related_works}. In Section~\ref{sec:non_acoustic_sim}, we discuss current non-acoustic language similarity approaches. Then, we describe our proposed acoustic language similarity approaches and their distinction from non-acoustic ones in Section ~\ref{sec:acoustic_sim}. Sections \ref{sec:lang_id}, \ref{sec:asr}, and \ref{sec:tts} contain our language family classification, speech recognition, and speech synthesis experiments, respectively. Finally, we summarize our results and propose future directions in Section~\ref{sec:concl}. All our code and other supplementary material can be found at \href{https://github.com/peter-yh-wu/cross-lingual}{https://github.com/peter-yh-wu/cross-lingual}.

\section{Cross-lingual Transfer}
\label{sec:related_works}


\subsection{Problem Statement}
\label{sec:problem_statement}

Given the set of all languages $S$, cross-lingual transfer generally refers utilizing $n$ source languages $\{s_1, \dots, s_n\} \subset S$ to improve task performance in target language $t \in S$, where $t$ is typically low-resource \cite{lin-graham-2019-choosing}. Assuming that we are using a data-driven approach, data $D_i$ would need to be chosen for each source language $s_i$. This definition reveals three challenging tasks: (1) choosing source languages $\{s_i\}$ from $S$; (2) choosing data $D_i$ for each $s_i$; (3) transferring the information in $\{D_1, \dots, D_n\}$ to the task in target language $t$. We focus on the \textbf{first} challenge in this work, and discuss its relation to other tasks below.

\subsection{Cross-lingual Techniques}

Many speech- and text-based works have approached the aforementioned third challenge of improving performance in target language $t$ using source data $\{D_i\}$ \cite{dalmia2018sequence, sercu2017network, toshniwal2018multilingual, pratap2020massively, hou2020large, chen2018multi, conneau2019unsupervised, conneau2020xlsr, chi2020cross}. To handle the second challenge of choosing the $D_i$'s, many works select all the source data from the same corpus \cite{conneau2020xlsr, conneau2019unsupervised} or an arbitrary set \cite{shinji17asr10, pratap2020massively, hou2020large}. Multiple works have approached the first challenge of choosing source languages $\{s_i\}$ in a similar manner \cite{conneau2020xlsr, conneau2019unsupervised, hou2020large, shinji17asr10, pratap2020massively}. These algorithms that utilize vast amounts of source language data are generally pre-training ones, and thus can be used in conjunction with our approach \cite{dalmia2018sequence, toshniwal2018multilingual, pratap2020massively, hou2020large, li2020towards}.

Many works have shown that certain subsets of source data are more suitable than others for cross-lingual transfer \cite{conneau2020xlsr, adams2019massively_shinji_asr}. At the dataset level, Li et al. \cite{li2019corpus} propose an approach to select suitable corpora by iteratively sampling from each dataset and gradually attending to less data using similarity scores. Since this work assumes a given initial set of candidate languages and datasets, it addresses the second challenge of selecting the data $\{D_i\}$. The output of our approach, which selects languages $\{s_i\}$, can thus be used as an input to theirs.

Within the text modality, Lin et al. explore using phylogenetic and typological features to select source languages \cite{lin-graham-2019-choosing}. Since speech data contains a wealth of additional information, we propose acoustic-based features and compare them with their features in speech recognition and speech synthesis. As discussed in Sections \ref{sec:lang_id}, \ref{sec:asr}, and \ref{sec:tts}, we observe unique advantages of our features in speech processing tasks. We describe all of these features and how we utilize them to measure language similarity below.

\subsection{Language Similarity}

In this work, we explore eight different language similarity approaches to select cross-lingual transfer languages for speech processing tasks. Namely, we focus on addressing the first challenge posed in Section \ref{sec:problem_statement} of choosing source languages $\{s_i\}$. All eight approaches are dataset-independent. They can be used in any cross-lingual transfer task as a compact look-up table without requiring additional computation. 
We proceed by detailing non-acoustic similarity approaches in Section \ref{sec:non_acoustic_sim} and our proposed acoustic ones in Section \ref{sec:acoustic_sim}.

\section{Non-Acoustic Language Similarity}
\label{sec:non_acoustic_sim}

\subsection{Non-Acoustic Approaches}

For our non-acoustic language similarity approaches, we use the six distance metrics from Lin et al. \cite{lin-graham-2019-choosing, littell-etal-2017-uriel}: genetic, inventory, syntactic, phonological, featural, and geographic. Sections \ref{sec:lang_fam_trees} and \ref{sec:phylo_typo_lang_sim} describe the background and details of the first five approaches. Since we observed that their geographic distance method would sometimes map languages to the same location, we utilize an additional geographic measure, as discussed in Section \ref{sec:geo}.

\subsection{Language Family Trees}
\label{sec:lang_fam_trees}

\begin{figure}[htb]
\centering
     \begin{minipage}[b]{0.30\textwidth}
         \centering
         \centerline{\includegraphics[width=\textwidth]{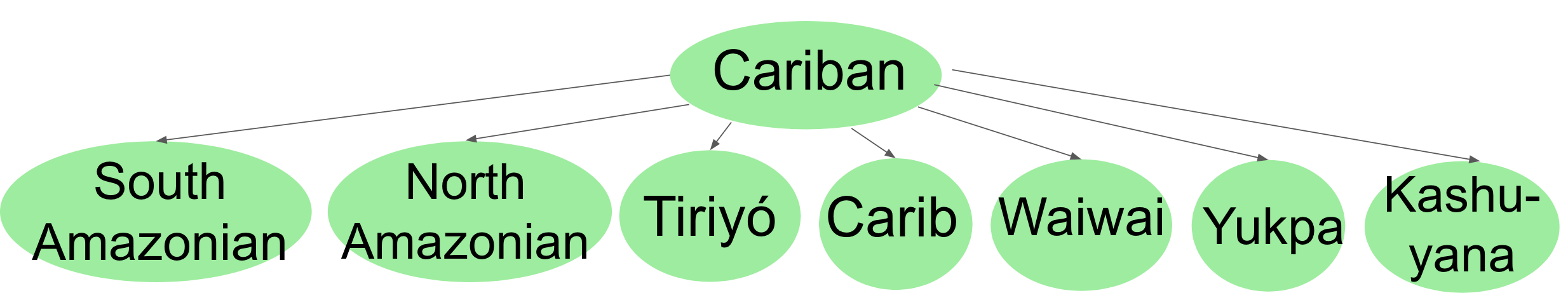}}
         \centerline{(a) Ethnologue}\medskip
         \label{fig:ethnologue}
     \end{minipage}
     \hfill
     \begin{minipage}[b]{0.13\textwidth}
         \centering
         \centerline{\includegraphics[width=\textwidth]{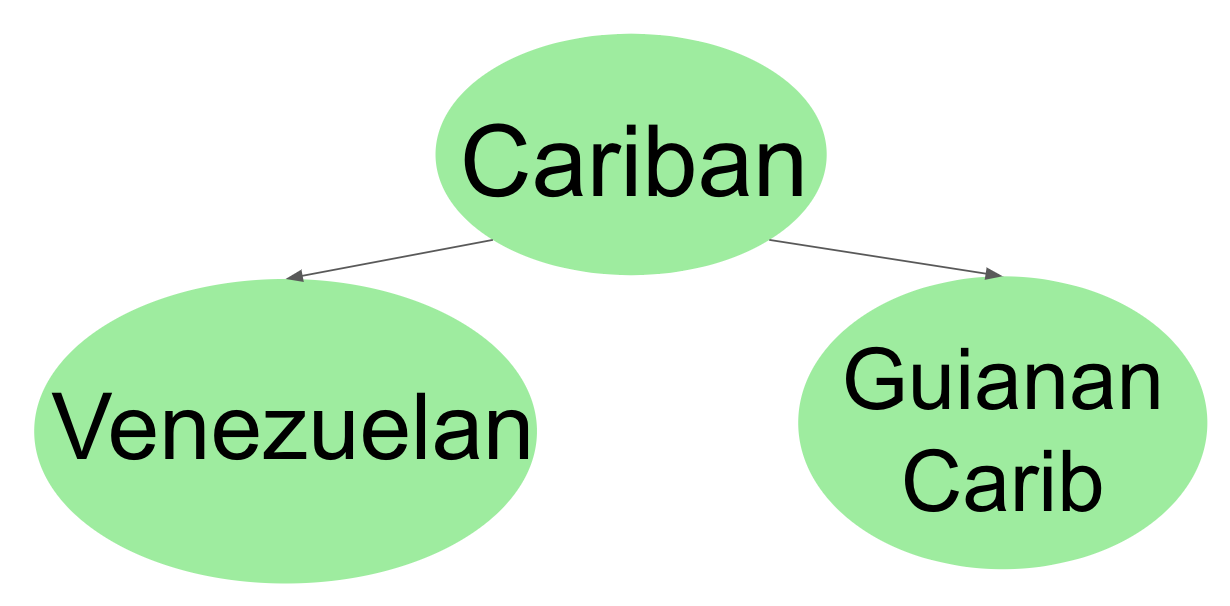}}
         \centerline{(b) Wikipedia}\medskip
         \label{fig:wikipedia}
     \end{minipage}
     \hfill
     \begin{minipage}[b]{0.4\textwidth}
         \centering
         \centerline{\includegraphics[width=\textwidth]{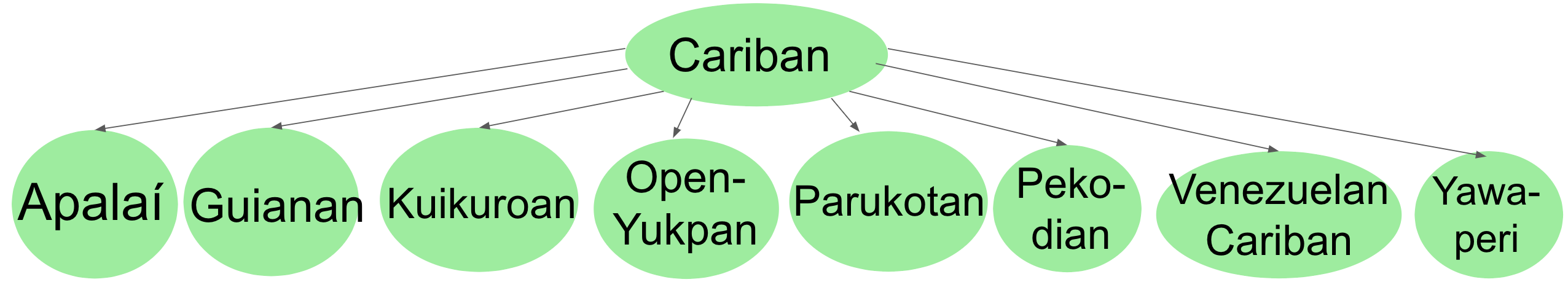}}
         \centerline{(c) Glottolog}\medskip
         \label{fig:glottolog}
     \end{minipage}
\caption{Top two levels of the Cariban language family tree based on three different linguistics data sources \cite{ethnologue, wu2020automatically, glottolog}.}
\label{fig:cariban_trees}
\vspace{-15px}
\end{figure}

Multiple sources have devised ways to categorize languages into families, using genealogical, typological, and other linguistic information \cite{ethnologue, glottolog, wals}. Several works categorize languages under a hierarchy of language families, which we can visualize as a directed acyclic graph where family nodes have languages and subfamilies as children. Since categorization decisions differ between linguists, these language family trees can vary noticeably between sources, as shown in Figure \ref{fig:cariban_trees}. The large number of different linguistic attributes thus offers us many different ways of defining language similarity, which we discuss below.

\subsection{Phylogenetic and Typological Distances}
\label{sec:phylo_typo_lang_sim}

As in Lin et al. and the URIEL typological database \cite{lin-graham-2019-choosing, littell-etal-2017-uriel}, our five language distance metrics that incorporate phylogenetic and typological information are:
\begin{enumerate}
  \item \textbf{Genetic Distance:} The distance between languages in Glottolog's world language family tree \cite{glottolog}. For reference, a subset of this tree is depicted in Figure~\ref{fig:cariban_trees} above.
  \item \textbf{Inventory Distance:} The cosine distance between binary vectors extracted from the PHOIBLE database \cite{phoible}. Each vector dimension has phonetic information about the presence of sounds \cite{littell-etal-2017-uriel}.
  \item \textbf{Syntactic Distance:} The cosine distance between syntax feature vectors utilizing information from the World Atlas of Language Structures (WALS) and Syntactic Structures of World Languages (SSWL) \cite{wals, sswl}.
  \item \textbf{Phonological Distance:} The cosine distance between vectors containing phonological information from Ethnologue and WALS \cite{ethnologue, wals}.
  \item \textbf{Featural Distance:} The cosine distance between vectors incorporating features from the above four \cite{littell-etal-2017-uriel}.
\end{enumerate}

For these five language similarity approaches, we use the pre-computed lang2vec distances provided in URIEL \cite{littell-etal-2017-uriel}. The genetic distance between any two languages is a real value in $[0,1]$, and all the aforementioned vectors are element-wise non-negative with norm greater than $0$. Thus, since cosine distance is given by
\begin{equation}
  d_{cos}(x,y) = 1-\frac{x\cdot y}{\left\lVert x\right\rVert \left\lVert y\right\rVert},
  \label{eq_cos}
\end{equation}
the other four distance functions have range $(0,1]$.

\subsection{Geographic Distance}
\label{sec:geo}

The sixth non-acoustic language similarity approach that we use is geographic distance, which we measure in two ways. Our first way is via the orthodromic distance approach described in Lin et al. and URIEL, using pre-computed lang2vec distances \cite{lin-graham-2019-choosing, littell-etal-2017-uriel}. We observed that languages may be mapped to the same location, and thus utilize an additional geographic distance measure. Namely, we obtain language coordinates using Wilderness metadata and compute the geodesic distance with the GeoPy library \cite{black2019cmu_wilderness, karney2013algorithms}.\footnote{\href{https://github.com/geopy/geopy}{https://github.com/geopy/geopy}} Unless stated otherwise, we report results in this work using the first approach. 

\section{Acoustic Language Similarity}
\label{sec:acoustic_sim}

\subsection{Acoustic Approaches}
\label{sec:acoustic}

We propose two acoustic-based approaches for measuring language similarity. In contrast to the non-acoustic approaches in Section \ref{sec:non_acoustic_sim}, these methods are entirely data driven and thus do not rely on expert linguistic knowledge. Additionally, they leverage information directly from speech data, offering opportunities to learn language properties suitable for downstream speech tasks. Our acoustic-based approaches are motivated by i-vector and x-vector language recognition works \cite{dehak2011language, snyder2018spoken}. While these works study tens of languages, we show that ours can successfully support hundreds. As far as we are aware, this work is also the first to compare speech vectors with other language similarity approaches and systematically study their uses in downstream speech tasks.

Generally, both involve training a neural network to classify languages given multilingual speech data. We generate a vector embedding for a language by averaging the outputs of a designated embedding layer across speech samples from that language. Then, we measure language distance as the distance between the language embeddings. The primary difference between our approaches is that one aligns its speech representations with another space, as detailed below.

Formally, our approaches are each defined by a model $M$ and a size-$N$ dataset $D = \{(x_n, y_n)\}_{n=1}^N$, where the $x$'s are model inputs and the $y$'s are language labels. $x$ is a speech representation in our first approach and a (speech, text) pair in our second. Let $M_e(x)$ be the output of the embedding layer, and $M_c(M_e(x))$ be the output of the language classification layer. In our first approach, we train $M$ on a subset $D_{train} \subset D$ using a classification loss $f_c(M_c(M_e(x)), y)$. 
Then, we generate an embedding $e_l$ for language $l$ using samples $D_l \subset \{ x_n | y_n = l \land (x_n, y_n) \in D\}$. We compute $e_l$ as the mean of the embedding layer outputs, or $\frac{1}{|D_l|}\sum_{x_n \in D_l} M_e(x_n)$. In this work, we measure language similarity using cosine distance, given in Equation \ref{eq_cos}. We refer to this approach as the \textbf{speech-based approach}.

For our second approach, we add a term to our loss function and assume each $x_n$ is a pair $(v_n, w_n)$ of speech and text data. Our new loss is given by $f(x, y) = f_c(M_c(M_e(v)), y) + f_a(M, x, y)$,
where $f_a$ is an alignment loss. In this work, we train a text encoder $T$ jointly with $M$ and define $f_a$ as
\begin{equation}
  f_a(M, x, y) = f_c(M_c(T(w)), y) + \alpha d_{cos}(M_e(v), T(w)),
  \label{eq_align}
\end{equation}
where $d_{cos}$ is the cosine distance function in Equation \ref{eq_cos} and $\alpha$ is a hyperparameter. We create language embeddings in a similar manner as the first approach, feeding $v$ into $M_e$ instead of $x$. We refer to this approach as the \textbf{multimodal-based approach} and discuss specific architectures used for both approaches in Section~\ref{sec:model}.


\subsection{Acoustic Dataset}
\label{sec:dataset}

\begin{figure}[htb]
  \centering
  \centerline{\includegraphics[width=\linewidth]{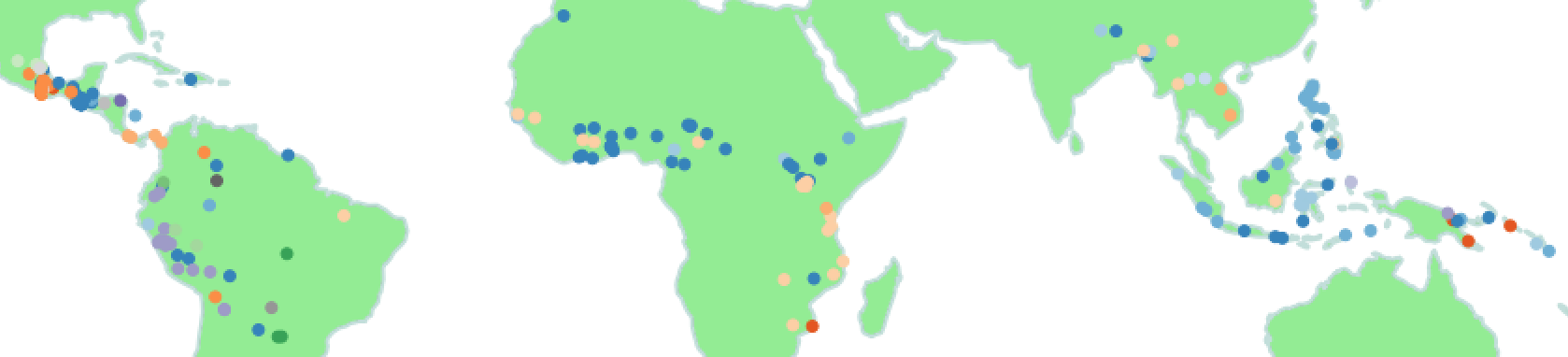}}
  \caption{Chosen Wilderness languages, colored by Ethnologue language family \cite{black2019cmu_wilderness, ethnologue}.}
  \label{fig:lang_map}
 \vspace{-15px}
\end{figure}

We use the Wilderness dataset to train the neural models in our acoustic similarity approaches from Section \ref{sec:acoustic} \cite{black2019cmu_wilderness}. This dataset is suitable for our acoustic approaches since it contains parallel speech and text data for over 700 languages. In this work, we focus on using data that are considered as being good or very good quality in Wilderness paper. Namely, we choose languages with data that could train a random forest Clustergen speech synthesizer with Mel-cepstral distortion (MCD) under a threshold $c$ \cite{black2006clustergen, black2015random, kominek2007voice, kubichek1993mel}. We let $c = 5.5$ since this is the middle value of the good quality range used in the Wilderness work \cite{black2019cmu_wilderness}. As discussed in that paper, MCD serves as an objective measure of synthesis quality, which in turn has been shown to correlate with alignment and dataset quality \cite{black2019cmu_wilderness, black2007statistical, voiers1977diagnostic}.
Among this subset of languages, we choose the low-resource ones, which we define as those with limited amounts of public data outside of Wilderness. This yields $195$ languages, with $22.8\pm 5.9$ hours of speech per language. Figure \ref{fig:lang_map} visualizes the location of these languages.

\subsection{Models for Acoustic Approaches}
\label{sec:model}

For both acoustic approaches proposed in Section \ref{sec:acoustic}, we use a convolutional network based on VGGVox as our speech encoder $M_e$, with an output dimension of $512$ \cite{voxceleb_vggvox} and two-dimensional adaptive max pooling. Each input to the speech encoder is an $80$-channel normalized Mel-scaled spectrogram with an FFT window length of $800$, hop size of $200$, window size of $800$, and sampling rate of $16000$ Hz. For all experiments, we use normalized embeddings as in Khosla et al. \cite{supcon2020khosla}, Adam optimization with learning rate $10^{-3}$, and batch size $128$. We explore two classification losses for our speech-based approach: cross-entropy and SupCon \cite{supcon2020khosla}, referred to as CE and SC in Tables \ref{tab:asr_spearman}, \ref{tab:finetune_asr}, and \ref{tab:tts_spearman}. Classifier $M_c$ is a ReLU activation followed by a fully connected (FC) layer with output dimension $195$ in the former, and an FC layer with 128 outputs in the latter. We replace the original SupCon augmentations with SpecAugment \cite{Park2019SpecAugmentAS}.

For the multimodal-based approach described in Section \ref{sec:acoustic}, we use an LSTM-based model as our text encoder $T$. Specifically, we feed each input into an embedding matrix with embedding dimension of $256$, followed by an LSTM with hidden dimension $128$. We then apply max pooling across the sequence dimension and feed the resulting vector into an FC layer with output dimension $128$. We chose max pooling instead of a more complex approach like attention, since the latter did not improve performance in the language family classification task in Section \ref{sec:lang_id}. To standardize text inputs across languages, we romanize them using UniTran, as described in the Wilderness paper  \cite{black2019cmu_wilderness, qian-etal-2010-unitran}. For all experiments with the multimodal-based approach, we set the alignment hyperparameter $\alpha$ to $3*10^{-2}$. Each model takes a few GPU days to train and a few GPU hours to extract all the language embeddings on an RTX 2080 Ti.

\begin{figure}[htb]
  \centering
  \centerline{\includegraphics[width=\linewidth]{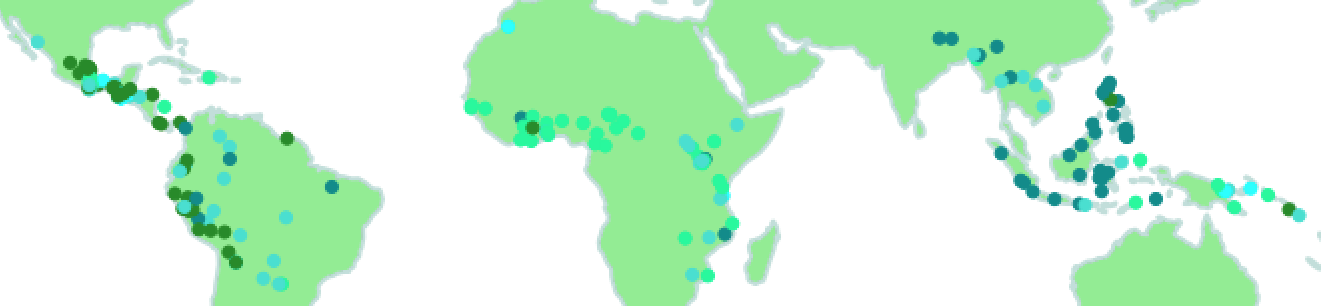}}
  \caption{$k$-means assignments of language embeddings extracted from speech data, where $k=5$. More details are in Section \ref{sec:model}.}
  \label{fig:k_means}
 \vspace{-15px}
\end{figure}

To check whether our language embeddings are interpretable, we compare their $k$-means clustering assignments with their geographic locations. Here, we train the classifier for the speech-based approach described in Section \ref{sec:acoustic} on a random $80\%$-$10\%$-$10\%$ train-val-test split of our $195$-language Wilderness subset using cross-entropy. For each language, we make an embedding using all of the test data in that language. Figure \ref{fig:k_means} colors the locations of these languages based on their $k$-means clustering assignments. Since assignments to the same cluster generally seem close on the map, it appears that our embeddings contain some degree of geographical information. Additionally, Figure \ref{fig:lang_map} and Figure \ref{fig:k_means} share many boundaries where language colorings change, suggesting that our embeddings capture some language family information as well.

\section{Language Family Classification}
\label{sec:lang_id}

In addition to computing language similarity, our acoustic approaches can also be used for classifying language families from speech. We explore this through a zero-shot task using our 195-language Wilderness subset from Section \ref{sec:dataset} and our two neural classifiers from Section \ref{sec:model}. More specifically, we separate the languages into a random $80\%$-$10\%$-$10\%$ train-validation-test split, yielding $156$, $19$, and $20$ languages. Then, we train our models to classify the $156$ training languages. During evaluation, we map the predicted language to its Ethnologue language family and compare whether the family matches that of the ground truth language. Validation and test accuracy are calculated over the $24$ Ethnologue language families containing our $195$ languages \cite{ethnologue}, and we select the best models based on validation accuracy. We also evaluate an untrained version of our Section \ref{sec:model} speech-based classifier, called the random baseline in Table \ref{tab:lang_id}.

The two trained models have noticeably higher test set accuracies than the random baseline, indicating that the models learned to perform zero-shot language family classification. Also, the multimodal-based approach outperforms the speech-based one, suggesting that the former encoded more linguistic content into its embedding space by aligning with textual representations. While speech-based classification has been widely explored \cite{wu2019ordinal, salesky2021sigtyp}, we believe that this is the first work to indicate its potential for classifying language families.

\begin{table}[tt]
  \caption{Language family classification from speech on unseen languages. More details are in Section \ref{sec:lang_id}.}
  \label{tab:lang_id}
  \centering
  \begin{tabular}{ l | c c}
    \toprule
    \multicolumn{1}{c}{\textbf{Approach\textbackslash Split}} & \multicolumn{1}{c}{\textbf{Val}} & \multicolumn{1}{c}{\textbf{Test}} \\
    \midrule
    Random & $0.00$ & $0.00$~~               \\
    Multimodal & $0.29$ & $0.22$~~       \\
    Speech & $0.25$ & $0.20$~~       \\
    \bottomrule
  \end{tabular}
  \vspace{-15px}
\end{table}

\section{Speech Recognition}
\label{sec:asr}

\subsection{ASR Task}
\label{sec:asr_task}

We also analyze the performance of our language similarity approaches in a cross-lingual speech recognition (ASR) task. Namely, we train monolingual ASR models on a set of source languages and evaluate them on a set of target languages. Thus, experiments using source languages different from target ones are zero-shot. For all ASR experiments, we romanize text using Unidecode\footnote{\href{https://pypi.org/project/Unidecode/}{https://pypi.org/project/Unidecode}} and evaluate performance using character error rate (CER). We note that the transliteration helps to make zero-shot ASR possible. To measure the suitability of each language similarity approach for cross-lingual ASR, we calculate the Spearman correlation between the language similarities and the CERs. Details on the specific sets of source-target language pairs used are described below. As mentioned in Section \ref{sec:problem_statement}, we focus 
here on how to choose the source languages rather than how to improve the subsequent steps in the cross-lingual transfer procedure.

Unless mentioned otherwise, we use an ESPNet Transformer ASR model comprised of a $12$-block Transformer encoder, a $6$-block Transformer decoder, and a CTC module \cite{watanabe2018espnet, karita2019espnet_transformer}. Additionally, we use a byte pair encoding vocabulary size of $1000$, decoding beam size of $10$, and an optimization procedure similar to that of Vaswani et al. \cite{vaswani2017attention}. Further details for reproducibility are provided in the accompanying code.

In addition to data from Wilderness \cite{black2019cmu_wilderness}, we also conduct ASR experiments using other public datasets. 
Namely, we use Javanese and Sundanese data from Kjartansson et al. \cite{java_sunda}, Iban data from Juan et al. \cite{juan2015iban}, and Indonesian, Hindi, Vietnamese, and Hakha Chin data from Common Voice Corpus 6.1 \cite{ardila2019commonvoicecv}. 
We chose these datasets since they are all public, their languages are relatively near each other, and most of these languages are low-resource.
Train-validation-test splits follow those defined in the respective works \cite{java_sunda, juan2015iban, ardila2019commonvoicecv}. We train our cross-lingual models using the source language training and validation sets and evaluate them on the target language test sets. Further details on how we use these datasets are described below.

\subsection{Motivation for Cross-lingual ASR}
\label{sec:asr_motivation}

Our first ASR experiment compares cross-lingual models with monolingual and multilingual ones, all evaluated on the Indonesian and Iban test sets described in Section \ref{sec:asr_task}. We train each monolingual model using data from the same dataset as the respective test set. Each cross-lingual model trains on a monolingual dataset in a language different from the test ones. Here, these monolingual datasets are the Javanese dataset by Kjartansson et al. \cite{java_sunda}, LibriSpeech \cite{panayotov2015librispeech}, and GigaSpeech \cite{chen2021gigaspeech}. Our two multilingual models are trained on the $8$-language MLS \cite{pratap2020mls} and the $52$-language version of Hou et al. \cite{hou2020large, watanabe2018espnet}. The training data for all models are the same as those defined by the original works. The monolingual and Javanese models are described in Section \ref{sec:asr_task}, and the rest are Conformers and ESPnet Transformers detailed in the ESPnet Model Zoo \cite{gulati2020conformer, watanabe2018espnet, karita2019espnet_transformer}.\footnote{\href{https://github.com/espnet/espnet_model_zoo}{https://github.com/espnet/espnet\_model\_zoo}} Table \ref{tab:id_cer} contains the CERs of all models on the two test sets.

By leveraging more training data, multiple models outperform the monolingual ones, which train on the training sets in the target datasets. Also, despite using less training data than multilingual and other cross-lingual approaches, Javanese performs the best on both test sets. Since Javanese is closer to the target languages than the other sources, these results suggest the usefulness of leveraging data from similar languages to build ASR systems for low-resource languages. The language similarity approaches in Sections \ref{sec:non_acoustic_sim} and \ref{sec:acoustic_sim} offer ways to automatically identify such similar languages.
For example, Javanese is closer to Indonesian than English for both our acoustic measures and the majority of the non-acoustic ones. Thus, we proceed to study the correlation between language similarity and performance in downstream speech tasks.

\begin{table}[th]
  \caption{ASR performance (CER) of monolingual, cross-lingual, and multilingual models evaluated on Indonesian and Iban test sets \cite{ardila2019commonvoicecv, juan2015iban}. Row $1$ contains the monolingual ASR CERs. For reference, we provide the amount of training data for each model. More details are in Section \ref{sec:asr_motivation}.}
  \label{tab:id_cer}
  \centering
  \begin{tabular}{ l | c c c }
    \toprule
    \multicolumn{1}{c}{\textbf{Training Data}} &
    \multicolumn{1}{c}{\textbf{Hrs.}} &
    \multicolumn{1}{c}{\textbf{Indo.}} & \multicolumn{1}{c}{\textbf{Iban}} \\
    \midrule
    Original Set \cite{ardila2019commonvoicecv, juan2015iban} & $< 10$ & $54.7$ & $75.7$~~             \\
    Javanese \cite{java_sunda} & $300$ & $\mathbf{35.4}$ & $\mathbf{50.4}$~~             \\
    LibriSpeech \cite{panayotov2015librispeech} & $1,000$ & $74.2$ & $56.3$~~ \\
    GigaSpeech \cite{chen2021gigaspeech} & $10,000$ & $55.3$ & $59.8$~~ \\
    MLS \cite{pratap2020mls} & $50,000$ & $70.5$ & $62.0$~~      \\
    OpenLI52 \cite{hou2020large} & $5,000$ & $41.8$ & $50.5$~~ \\
    \bottomrule
  \end{tabular}
  \vspace{-15px}
\end{table}

\subsection{Wilderness Zero-Shot ASR}
\label{sec:wld_asr}

To compare the language similarity methods in Sections \ref{sec:non_acoustic_sim} and \ref{sec:acoustic_sim}, we first perform zero-shot ASR using Javanese and Sundanese as our source languages and Wilderness languages as our targets. Here, we randomly select ten languages from our $195$-language Wilderness subset and evaluate each language using all of its data.\footnote{Wilderness ASR $6$-letter codes are ACCIBS, AGUNVS, BOATBL, CSOTBL, FRDWBT, JACWBT, NODWBT, POHPOC, STNBSP, and VUNBST.} Table \ref{tab:asr_spearman} contains the Spearman correlations between ASR performance (CER) and different similarity measures, where each number uses a single source language and all the targets. Our multimodal-based similarity approach performs the best overall, followed by the genetic, speech-based, and geographic measures. Also, our speech-based approaches perform about the same. This suggests that cross-entropy may be sufficient to train our acoustic measure, potentially since its training dataset is large and consists of non-noisy labels. Another benefit of cross-entropy is its low computational costs during training relative to objectives like contrastive loss. To explore the effects of ensembling, we combined the best linguistic language similarity measure for each source language with our best acoustic one by rescaling their language similarities to $[0, 1]$ and then averaging the rescaled values. This combination performs the best overall, suggesting the usefulness of leveraging both non-acoustic and acoustic information. We note that both our speech-based approach and the target languages here use the $195$-language Wilderness subset described in Section \ref{sec:dataset}. Thus, in the next section, we study how well our similarity measure generalizes to cases where both the source and target language data are not from Wilderness.

\begin{table}[tt]
  \caption{Spearman correlations between language similarity measures and cross-lingual ASR performance (CER), using Javanese and Sundanese as source languages. More details about Wilderness (Wild.) and non-Wilderness (Non-W.) experiments are in Sections \ref{sec:wld_asr} and \ref{sec:non_wld_asr}.}
  \label{tab:asr_spearman}
  \centering
  \begin{tabular}{ l | c c | c c }
    \toprule
    \multicolumn{1}{c}{\textbf{}} & \multicolumn{2}{c}{\textbf{Javanese}} & \multicolumn{2}{c}{\textbf{Sundanese}} \\
    \multicolumn{1}{c}{\textbf{Similarity}} & \multicolumn{1}{c}{\textbf{Wild.}} & \multicolumn{1}{c}{\textbf{Non-W.}} & \multicolumn{1}{c}{\textbf{Wild.}} & \multicolumn{1}{c}{\textbf{Non-W.}} \\
    \midrule
    Syntactic & $-0.70$ & $0.52$ & $0.12$ & $0.14$~~               \\
    Geographic & $0.65$ & $0.68$ & $0.29$ & $0.72$~~       \\
    Phonological & $0.00$ & $0.04$ & $-0.08$ & $0.31$~~       \\
    Genetic & $0.70$ & $\mathbf{0.81}$ & $\mathbf{0.52}$ & $0.45$~~               \\
    Inventory & $-0.25$ & $0.43$ & $0.04$ & $0.39$~~       \\
    Featural & $-0.35$ & $0.54$ & $0.12$ & $0.22$~~       \\
    Multimodal & $\mathbf{0.85}$ & $0.43$ & $0.44$ & $0.29$~~       \\
    Speech (SC) & $0.71$ & $0.36$ & $0.25$ & $0.64$~~ \\
    Speech (CE) & $0.73$ & $0.64$ & $0.27$ & $\mathbf{0.79}$~~ \\
    \midrule
    Ensemble & $\mathbf{0.93}$ & $0.79$ & $0.50$ & $0.75$~~ \\
    \bottomrule
  \end{tabular}
  \vspace{-15px}
\end{table}

\subsection{Non-Wilderness Zero-Shot ASR}
\label{sec:non_wld_asr}

To study the generalizability of our speech-based similarity measure, we perform cross-lingual ASR using source and target language data that are both not from Wilderness. Namely, our source languages are Javanese and Sundanese, and our target languages are these two plus Iban, Indonesian, Hindi, Vietnamese, and Hakha Chin \cite{java_sunda, juan2015iban, ardila2019commonvoicecv}. Section \ref{sec:asr_task} provides more information about these data. Table \ref{tab:asr_spearman} contains the Spearman correlations between ASR performance (CER) and different language similarity measures, where each number uses a single source language and all the targets. The cross-entropy speech-based approach performs the best overall, followed by the geographic, genetic, and SupCon speech-based measures. These approaches also performed well in the Wilderness ASR task discussed in Section \ref{sec:wld_asr}. Moreover, ensembling in the same manner as that task also yields the highest average correlation, suggesting that these trends may hold even in non-Wilderness data settings.

\subsection{ASR with Fine-tuning}
\label{sec:finetune_asr}

We also explore the performance of our similarity measures with fine-tuned ASR models. For each source-target pair in our zero-shot experiments above, we fine-tuned the ASR model on a random ten-minute subset of the target language dataset that is disjoint from the evaluation data. We again observe that the speech-based, multimodal-based, geographic, and genetic measures perform the best, as well as the ensemble having the highest mean correlation.

\begin{table}[tt]
  \caption{Spearman correlations between language similarity measures and fine-tuned cross-lingual ASR performance (CER), using Javanese and Sundanese as source languages. More details are in Section \ref{sec:finetune_asr}.}
  \label{tab:finetune_asr}
  \centering
  \begin{tabular}{ l | c c | c c }
    \toprule
    \multicolumn{1}{c}{\textbf{}} & \multicolumn{2}{c}{\textbf{Javanese}} & \multicolumn{2}{c}{\textbf{Sundanese}} \\
    \multicolumn{1}{c}{\textbf{Similarity}} & \multicolumn{1}{c}{\textbf{Wild.}} & \multicolumn{1}{c}{\textbf{Non-W.}} & \multicolumn{1}{c}{\textbf{Wild.}} & \multicolumn{1}{c}{\textbf{Non-W.}} \\
    \midrule
    Syntactic & $-0.45$ & $0.48$ & $0.26$ & $-0.04$~~               \\
    Geographic & $0.77$ & $0.42$ & $0.65$ & $0.72$~~       \\
    Phonological & $-0.53$ & $0.34$ & $-0.17$ & $0.11$~~       \\
    Genetic & $0.66$ & $0.58$ & $0.65$ & $0.34$~~               \\
    Inventory & $-0.60$ & $0.21$ & $-0.07$ & $0.29$~~       \\
    Featural & $-0.67$ & $0.42$ & $-0.32$ & $0.04$~~       \\
    Multimodal & $\mathbf{0.93}$ & $\mathbf{0.64}$ & $\mathbf{0.77}$ & $0.18$~~       \\
    Speech (SC) & $0.84$ & $0.50$ & $0.54$ & $0.61$~~ \\
    Speech (CE) & $0.84$ & $0.43$ & $0.43$ & $\mathbf{0.75}$~~ \\
    \midrule
    Ensemble & $0.90$ & $\mathbf{0.64}$ & $\mathbf{0.77}$ & $0.71$~~ \\
    \bottomrule
  \end{tabular}
  \vspace{-15px}
\end{table}

\section{Speech Synthesis}
\label{sec:tts}

\subsection{TTS Task}

We also analyze our similarity measures in cross-lingual text-to-speech (TTS) tasks. Namely, we train monolingual TTS models on a set on source languages and evaluate them on a set of target languages. Thus, experiments using source languages different from target ones are zero-shot. Section \ref{sec:wld_stat_tts} describes our experiments using a statistical model with Wilderness data, and Section \ref{sec:non_wld_neural_tts} describes our experiments using a neural model with non-Wilderness data. Similarly to our ASR experiments in Section \ref{sec:asr}, we romanize text using Unidecode, which helps make zero-shot TTS possible. Details on evaluation metrics and the specific sets of source-target language pairs used are described below. As mentioned in Section \ref{sec:problem_statement}, we focus 
here on how to choose the source languages rather than how to improve the subsequent steps in the cross-lingual transfer procedure.

\begin{table}[th]
  \caption{Spearman correlations between language similarity methods and TTS performance, measured with MCD for the Wilderness experiment and MOS for the non-Wilderness one. A more positive correlation with MCD is better, and a more negative correlation with MOS is better. More details are in Sections \ref{sec:wld_stat_tts} and \ref{sec:non_wld_neural_tts}.}
  \label{tab:tts_spearman}
  \centering
  \begin{tabular}{ l | c | c c }
    \toprule
    \multicolumn{1}{c}{\textbf{}} & 
    \multicolumn{1}{c}{\textbf{Sim. vs. MCD} \textcolor{ForestGreen}{$\uparrow$}} &
    \multicolumn{2}{c}{\textbf{Sim. vs. MOS} \textcolor{ForestGreen}{$\downarrow$}} \\
    \multicolumn{1}{c}{\textbf{Similarity}} &
    \multicolumn{1}{c}{\textbf{}} &
    \multicolumn{1}{c}{\textbf{Hindi}} &
    \multicolumn{1}{c}{\textbf{Telugu}} \\
    \midrule
    Syntactic & $0.32$ & $-0.76$ & $-0.30$~~               \\
    Geographic & $0.27$ & $-0.82$ & $-0.50$~~       \\
    Phonological & $-0.03$ & $-0.65$ & $-0.95$~~       \\
    Genetic & $0.40$ & $-0.80$ & $-0.53$~~               \\
    Inventory & $0.12$ & $-0.67$ & $-0.70$~~       \\
    Featural & $0.33$ & $-0.65$ & $-0.71$~~       \\
    Multimodal & $0.50$ & $\mathbf{-0.87}$ & $\mathbf{-1.0}$~~       \\
    Speech (SC) & $\mathbf{0.53}$ & $\mathbf{-0.87}$ & $\mathbf{-1.0}$~~       \\
    Speech (CE) & $0.47$ & $\mathbf{-0.87}$ & $-0.90$~~       \\
    \midrule
    Ensemble & $0.52$ & $\mathbf{-0.97}$ & $\mathbf{-1.0}$~~       \\
    \bottomrule
  \end{tabular}
  \vspace{-15px}
\end{table}

\begin{table*}[t]
  \caption{MOS ratings of cross-lingual Indic TTS models, where Hindi and Telugu are our target languages. We also provide the distances calculated by our top six language similarity measures from Table \ref{tab:tts_spearman}. More details are in Section \ref{sec:non_wld_neural_tts}.
  }
  \label{tab:mos}
  \centering
  \begin{tabular}{ l | c c c c c c c | c c c c c c c }
    \toprule
    \multicolumn{1}{c}{\textbf{}} & \multicolumn{5}{c}{\textbf{Hindi}} &
    \multicolumn{5}{c}{\textbf{Telugu}} \\
    \multicolumn{1}{c}{\textbf{Source}} &
    \multicolumn{1}{c}{\textbf{MOS}} &
    \multicolumn{1}{c}{\textbf{SC}} &
    \multicolumn{1}{c}{\textbf{Mul.}} &
    \multicolumn{1}{c}{\textbf{CE}} &
    \multicolumn{1}{c}{\textbf{Pho.}} &
    \multicolumn{1}{c}{\textbf{Inv.}} &
    \multicolumn{1}{c}{\textbf{Fea.}} &
    \multicolumn{1}{c}{\textbf{MOS}} &
    \multicolumn{1}{c}{\textbf{SC}} &
    \multicolumn{1}{c}{\textbf{Mul.}} &
    \multicolumn{1}{c}{\textbf{CE}} &
    \multicolumn{1}{c}{\textbf{Pho.}} &
    \multicolumn{1}{c}{\textbf{Inv.}} &
    \multicolumn{1}{c}{\textbf{Fea.}} \\
    \midrule
    Hindi & $\mathbf{4.8 \pm 0.4}$ & $0.00$ & $0.00$ & $0.00$ & $0.00$ & $0.00$ & $0.00$ & $3.4 \pm 0.9$ & $0.43$ & $0.57$ & $0.33$ & $0.30$ & $0.31$ & $0.40$~~       \\
    Kannada & $2.8 \pm 1.1$ & $0.05$ & $0.10$ & $0.07$ & $0.30$ & $0.44$ & $0.40$ & $4.2 \pm 0.7$ & $0.42$ & $0.56$ & $0.28$ & $0.00$ & $0.36$ & $0.40$~~       \\
    Marathi & $2.8 \pm 1.5$ & $0.12$ & $0.11$ & $0.14$ & $0.59$ & $0.40$ & $0.50$ & $2.9 \pm 0.9$ & $0.55$ & $0.65$ & $0.33$ & $0.64$& $0.36$ & $0.40$~~               \\
    Tamil & $1.5 \pm 0.5$ & $0.15$ & $0.24$ & $0.23$ & $0.59$ & $0.47$ & $0.50$ & $1.9 \pm 0.3$ & $0.58$ & $0.83$ & $0.41$ & $0.64$ & $0.43$ & $0.40$~~       \\
    Telugu & $2.1 \pm 1.2$ & $0.43$ & $0.57$ & $0.33$ & $0.30$ & $0.31$ & $0.40$ & $\mathbf{4.8 \pm 0.4}$ & $0.00$ & $0.00$ & $0.00$ & $0.00$ & $0.00$ & $0.00$~~       \\
    \bottomrule
  \end{tabular}
  \vspace{-15px}
\end{table*}

\subsection{Wilderness Statistical Speech Synthesis}
\label{sec:wld_stat_tts}

We first check the suitability of our similarity measures for cross-lingual TTS with Wilderness data \cite{black2019cmu_wilderness}. Namely, we randomly select $15$ languages from our $195$-language Wilderness subset described in Section \ref{sec:dataset} and use all $15$ as both sources and targets, yielding $225$ source-target language pairs.\footnote{Wilderness TTS $6$-letter codes are APRWBT, JACWBT, KEKSBG, KJBSBG, MAKLAI, MOPWBT, NASPNG, NPLWYI, POHPOC, QEJLLB, QULSBB, QWHLLB, TZBSBM, TZCSBM, and VUNBST.} We found that the data in each language was insufficient to build a neural TTS model \cite{wang2017tacotron}, and thus instead used the random forest Clustergen speech synthesizer as our TTS model \cite{black2006clustergen, black2015random}. Our training steps are the same as those in the Wilderness paper, and we evaluate models on the first chapter texts in each target language  \cite{black2019cmu_wilderness}. 
As discussed in Section \ref{sec:dataset}, we measure TTS performance objectively using Mel-cepstral distortion (MCD).

Table \ref{tab:tts_spearman} shows the Spearman correlations between TTS performance (MCD) and the language similarity measures. For each language similarity approach, we calculate the mean of $15$ Spearman correlations, each using one source and all the targets as in Section \ref{sec:asr}. Our acoustic approaches described in Section \ref{sec:acoustic} perform the best, followed by the genetic distance measure. We note that the trends observed here are similar to those from our ASR tasks in Section \ref{sec:asr}.

\subsection{Non-Wilderness Neural Speech Synthesis}
\label{sec:non_wld_neural_tts}

\paragraph{Non-Wilderness TTS Data}: In Section \ref{sec:wld_stat_tts} above, both our speech-based approach and the TTS data use our 195-language Wilderness subset described in Section \ref{sec:dataset}. Thus, we also study how well our similarity measure generalizes to cases where the TTS data are not from Wilderness. Namely, we train five monolingual neural TTS models on different Indic source languages and evaluate them on two Indic target languages. Our source languages are Hindi, Kannada, Marathi, Tamil, and Telugu, and our target languages are Hindi and Telugu, all from the CMU INDIC corpus \cite{wilkinson2016cmuindic}. For each language, we sort the utterances by file name and let the last $100$ be the test set, the $100$ before that be the validation set, and the remaining be the training set.

\paragraph{Non-Wilderness TTS Model}: Our models all use the same Transformer-TTS architecture, initialized using pre-trained weights from an English TTS model \cite{hayashi2020espnettts, li2019transformertts}. 
Namely, we use $4$ attention heads, $6$ layers and $1536$ units in both our encoder and decoder, and an optimization procedure similar to that of Vaswani et al. \cite{vaswani2017attention}.
We train models using the train and validation sets of source languages and evaluate synthesized texts in the target languages' test sets. TTS samples and further details for reproducibility are provided in the accompanying GitHub.

\paragraph{Non-Wilderness TTS Evaluation}: For each target language, we measure TTS performance using a MOS naturalness test performed by four listeners who are native speakers of that language. Namely, we ask evaluators to rate from $1$ to $5$ how much each sample sounds like natural speech in the target language, where $5$ is the highest. Our MOS evaluation data contains two samples from each source-target pair, randomly chosen from the synthesized test set samples. Listeners rate all the samples that have their target languages, yielding $40$ total MOS ratings per target.

Table \ref{tab:mos} summarizes these MOS test results, including the distances calculated by our top six language similarity measures from Table \ref{tab:asr_spearman}. Evaluators rated TTS samples with the same source and target the highest, as expected. Our proposed speech-based language similarity approach correctly identified the top two similar but different source languages for both targets. We compare this approach with others below.

\paragraph{Non-Wilderness TTS Correlations}: Table \ref{tab:tts_spearman} contains the Spearman correlations between the MOS values and different similarity measures, where each number uses a single target language and all the sources. Since the first geographic similarity measure from Section \ref{sec:geo} considers these languages to be the same location, we use the second one instead. We observe that our acoustic approaches yield the strongest correlations, as with our Wilderness TTS experiment in Section \ref{sec:wld_stat_tts}. Compared to our previous tasks, the ensemble again performs the best overall, but the genetic and geographic measures did not rank as high. This further suggests that language similarity measures grounded in speech are suitable for acoustic cross-lingual transfer.

\section{Conclusion and Future Directions}
\label{sec:concl}

In this work, we study how to choose source languages for acoustic cross-lingual transfer tasks through exploring eight different language similarity measures. Our proposed acoustic approaches often outperform other techniques in downstream tasks like speech recognition and text-to-speech tasks. This same approach can also be used to classify the families of unseen languages. Moving forward, we plan to train our acoustic approaches on all of the Wilderness data in order to strengthen their performance on languages outside our $195$-language subset \cite{black2019cmu_wilderness}. Our work can also be extended to a range of new acoustic cross-lingual transfer directions. These include leveraging language similarity in tasks like speech translation \cite{Wang2020ImprovingCT} and voice conversion \cite{wu2021vc}, as well as integrating cross-domain \cite{li2019corpus, gupta-etal-2021-task} and multimodal learning \cite{liang2020cross, liang2021multibench} insights into our methods.



\section{Acknowledgements}

We thank Sai Krishna Rallabandi and our listening test participants for helping us collect MOS values for our Indic TTS experiments. 
This work used the Extreme Science and Engineering Discovery Environment (XSEDE) \cite{towns2014xsede}, which is supported by National Science Foundation grant number ACI-1548562. Specifically, it used the Bridges system \cite{nystrom2015bridges}, which is supported by NSF award number ACI-1445606, at the Pittsburgh Supercomputing Center (PSC).

\bibliographystyle{IEEEbib}
\bibliography{strings,refs}

\end{document}